\documentclass[prb,preprint]{revtex4}

\usepackage{graphicx}

\begin{document}

\title{Dendritic flux penetration in Pb films with a periodic array of antidots}

\author{M.~Menghini, R.~J.~Wijngaarden}
\affiliation{Department of Physics and Astronomy, Faculty of
Sciences, Vrije Universiteit, De Boelelaan 1081, 1081HV Amsterdam,
The Netherlands}

\author{A.~V.~Silhanek\footnote{present address: MST-NHMFL, MS E536, Los Alamos National Laboratory, Los Alamos, NM 87544, USA.}}
\affiliation{Nanoscale Superconductivity and Magnetism Group,
Laboratory for Solid State Physics and Magnetism K. U. Leuven\\
Celestijnenlaan 200 D, B-3001 Leuven, Belgium}

\author{S. Raedts}
\affiliation{Nanoscale Superconductivity and Magnetism Group,
Laboratory for Solid State Physics and Magnetism K. U. Leuven\\
Celestijnenlaan 200 D, B-3001 Leuven, Belgium}

\author{V.~V.~Moshchalkov}
\affiliation{Nanoscale Superconductivity and Magnetism Group,
Laboratory for Solid State Physics and Magnetism K. U. Leuven\\
Celestijnenlaan 200 D, B-3001 Leuven, Belgium}

\date{\today}

\begin{abstract}

We explore the flux-jump regime in type-II Pb thin films with a
periodic array of antidots by means of magneto-optical
measurements. A direct visualization of the magnetic flux
distribution allows to identify a rich morphology of flux
penetration patterns. We determine the phase boundary $H^*(T)$
between  dendritic penetration at low temperatures and a smooth
flux invasion at high temperatures and fields. For the whole range
of fields and temperatures studied, guided vortex motion along the
principal axes of the square pinning array is clearly observed. In
particular, the branching process of the dendrite expansion is
fully governed by the underlying pinning topology. A comparative
study between macroscopic techniques and direct local
visualization shed light onto the puzzling $T-$ and
$H-$independent magnetic response observed at low temperatures and
fields. Finally, we find that the distribution of avalanche sizes
at low temperatures  can be described by a power law with exponent
$\tau \sim 0.9(1)$.

\end{abstract}

\pacs{PACS numbers: 74.76.Db, 74.60.Ge, 74.25.Dw,
74.60.Jg,74.25.Fy}

\maketitle

\section{Introduction}

Flux penetration in a type-II superconductor in the mixed state is
usually described by the Bean critical state model. In this
approximation it is assumed that a  balance between the pinning
force and the external magnetic pressure leads to a constant flux
gradient.\cite{bean} Similarly to a sand pile, this vortex
distribution is  metastable  and therefore it is bound to decay to
a lower energy configuration. The dynamic evolution towards the
equilibrium state is generally described as  flux creep  where
thermal or quantum fluctuations are needed to overcome a current
dependent pinning barrier $U(j)$.\cite{anderson} If  this process
takes place under isothermal conditions the creep is logarithmic
in time  and the field penetration is smooth with a flat flux
front.\cite{forkl,koblishka} In contrast, if the process is
perfectly adiabatic, the heat dissipation $\delta Q$ produced by
the vortex motion will give rise to a local increase of the
temperature $\delta T=\delta Q / C$, where $C$ is the specific
heat of the superconducting material. Since typically $dJ_c/dT<0$,
this local rise of temperature implies a reduction of the critical
current which in turn promotes further vortex motion thus yielding
a vortex avalanche. In this scenario, the  field penetration is
abrupt, giving rise to jumps in the magnetization,  and  develops
much faster than the creep relaxation process. These avalanches
(or flux-jumps) occur at low temperatures where critical currents
are high and the specific heat is small thus severely undermining
the potential technological applications of superconducting
materials.\cite{mints-review}

In most cases, flux penetration experiments are performed in thin
superconducting materials of slab geometry exposed to a field
perpendicular to the sample plane. It has been observed for many
samples that in this configuration the field penetrates via highly
branched expansions giving rise to a dendritic pattern of flux
channels.\cite{duran,johansen} Theoretical studies as well as
numerical simulations \cite{aranson,johansen}, have reproduced the
observed flux penetration patterns in thin films giving support to
a thermo-magnetic origin for these type of instabilities.

On the other hand, the influence of the pinning landscape on the
morphology and local characteristics of dendritic flux penetration
has not been yet fully addressed. A model system to investigate
this effect can be realized by tailoring a periodic pinning array
in superconducting samples. Vortex avalanches have been previously
detected \cite{terentiev,hebert,silhanek,vlasko-vlasov} in thin
films with  periodic pinning arrays. Global magnetization
measurements\cite{terentiev,hebert,silhanek} in samples with
arrays of antidots have shown that the region in the $H-T$ phase
diagram dominated by flux-jumps is more extended   as compared to
the case of plain films. Besides that, commensurability of
flux-jumps with the matching field of the pinning lattice and
invariance of the magnetization and flux-jump size distribution at
low temperatures and fields were observed. Moreover,
magneto-optical (MO) imaging \cite{vlasko-vlasov} in Nb patterned
films has shown that vortex avalanches along the principal
directions of the pinning lattice take place in zero-field cooling
(ZFC) experiments. From the theoretical point of view, Aranson
{\it et al.}\cite{aranson} have shown that a periodic spatial
modulation of the critical current gives rise to a branching
pattern of local temperature following the symmetry of the
underlying pinning array.

In this work we study the flux-jump regime in Pb thin films with
periodic pinning by means of MO. In order to separate and clearly
identify the effects of the engineered pinning potential similar
experiments were performed on plain Pb films.  The characteristics
of the samples and the MO technique are described in the following
section. Subsequently, the different types of flux penetration in
ZFC experiments at different temperatures are described.
Additionally,  phase boundary separating dendritic from smooth
penetration was determined. Finally, we present an analysis of the
evolution of dendrites with field and we study the avalanche size
distribution as a function of temperature.

\section{Samples and Experimental Procedure}

The experiments were conducted on Pb thin films with a square
array of antidots. The dimensions and critical temperature for
each sample are summarized in Table \ref{table}. In all patterned
samples the square antidot array consists of square holes with
lateral dimension $b = 0.8~\mu$m and period $d=1.5~\mu$m which
corresponds to a first matching field $\mu_0 H_1=0.92$ mT.
Simultaneously with each patterned film we deposited also an
unpatterned reference film on a SiO$_2$ substrate which allows us
to perform a direct comparison in order to ascertain the effects
of the pinning array (see Table I). Due to geometrical
characteristics these Pb thin films are type-II superconductors.
\cite{dolan,rodewald} From the temperature dependence of the upper
critical field $H_{c2}(T)$ of the plain films we have estimated a
superconducting coherence length $\xi(0) = 33 \pm 3$ nm. A more
detailed description of the sample preparation can be found in
Ref.[\onlinecite{sophie-crete}].

\begin{table}[ht]
\centering \caption{Lateral dimensions ($w_1$ and $w_2$),
thickness ($t$), and critical temperatures ($T_c$) for all the
films studied. AD indicates a square array of antidots and PF a
plain film.}

\begin{tabular}[b]{l c c c c}
\\
Sample & $w_1$ (mm) & $w_2$ (mm) & $t$ (nm) & $T_c$ (K)\\
     \hline

AD75 & 1.6 & 2.9 & 75 & 7.21 \\
AD65 & 2.3 & 2.5 & 65 & 7.21 \\
AD15 & 1.9 & 2.0 & 13.5 & 7.10 \\
PF15 & 2.2 & 3.1 & 13.5 & 7.10 \\
\label{table}
\end{tabular}
\end{table}

The local magnetic induction, $B$, just above the surface of the
sample was measured using a magneto-optical image lock-in
amplifier technique as described in Ref.[\onlinecite{MO}]. The
magnetic induction was sensed using an  indicator with in-plane
magnetization and large Faraday effect mounted on top of the
sample. The sample together with the indicator were mounted in a
specially designed cryogenic polarization microscope. The
experiments were performed in a commercial Oxford Instruments 7 T
vector magnet system.

\section{Results and Discussion}

\subsection{Dendrite morphology}

Magneto-optical imaging of Pb films with a square array of
antidots shows a rich variety of magnetic flux penetration in ZFC
experiments as a function of temperature. Fig.\,\ref{images}
summarizes the different morphologies of flux penetration observed
in these samples. The brighter regions correspond to high magnetic
fields while the dark ones indicate zero field. In the bottom part
of Figs.\,\ref{images} (a)-(d) magnetic domains from the magneto
optical garnet show up as a saw-tooth-like boundary between
regions with different contrast. These domains do not seem to
influence the flux pattern inside the sample and are irrelevant
for the discussion below. At low $T$ and $H$ finger-like dendrites
elongated in the direction perpendicular to the sample's border
are formed (Fig.\,\ref{images}(a)). As the temperature increases
to 5.5 K the dendrites become considerably larger and more
branched (see Fig.\,\ref{images}(b)). In the range, $5.5 < T \leq
6$ K, the magnetic field first penetrates smoothly up to
approximately $1/4$ of the sample width ($\mu_0H \sim 1.5$ mT) and
then suddenly a highly branched dendrite is formed. An example of
this behavior, also predicted theoretically \cite{aranson}, is
shown in Fig.\,\ref{images}(c). In the present sample we found
highly branched (tree-like) dendrites for  applied fields  up to 3
mT, for higher fields the penetration becomes uniform. It is
 noteworthily that in the finger-like regime the maximum length of
dendrites is limited by  the  half width of the sample whereas in
the region of highly-branched or tree-like dendrites, vortices can
extend much further into the sample. Finally, for $T>6\,$K a
smooth flux penetration is observed (Fig.\,\ref{images}(d)) in the
whole range of fields investigated and a Bean-like pattern
develops.

Within the regime dominated by avalanches it is found that the
main core of the dendrites and their ramifications are oriented
along the principal directions of the square array of antidots.
However, the influence of the underlying periodic pinning array is
not constrained to the flux-jump regime but can also be seen in
the smooth Bean-like penetration pattern. Indeed, a closer look at
the flux front for $T>6\,$K shows clear streaks aligned with the
pinning array as a result of preferential or guided motion of
vortices.\cite{pannetier} This result is consistent with previous
reports in low temperature as well as
 high temperature superconductor thin films with
periodic pinning.\cite{pannetier,radu,marco} On the other hand, it
has been theoretically shown that, in samples with random
disorder, during the initial ramping of the field hot magnetic
filaments propagate from the border of the sample. This, could
eventually also lead to streaks in the field
penetration.\cite{aranson} However, since we have observed a
filamentary penetration only in the patterned sample and not in
the plain film, we can rule out this possibility and attribute the
observed effect to the periodic pinning potential.

The influence of the square lattice of antidots on the flux
penetration becomes more evident when comparing the previous
results with those obtained in Pb plain films (PF)
(Figs.\,\ref{images} (e) and (f)). In this case, ZFC MO
experiments show that the vortex avalanche regime is constrained
to a smaller region of the $H-T$ phase diagram (see
Fig.\,\ref{hstar}(b)). Besides that, as can be clearly seen in the
flux patterns formed at low temperatures (Fig.\,\ref{images}(e)),
the morphology of the dendrites is quite different from the one
described above for antidot samples. In PF we observe that the
magnetic field bursts in highly disordered dendrites with no
particular orientation (other than the average imposed by the
screening currents) and with no characteristic size. These
features are similar to those previously reported for Nb and
MgB$_2$ plain films.\cite{duran,johansen} Finally,  a smooth
penetration is found at high temperatures and fields, as in the
case of the patterned sample.

In all cases, we have observed that the dendrites develop rather
abruptly, $v > 10$ m/s, according to the limit imposed by our
experimental temporal resolution. Previously, it was shown that
this velocity can indeed be much higher. \cite{leiderer} Besides
this, dendrites nucleate at the edge of the sample in random
positions which do not reproduce if the experiment is repeated.
This indicates that their appearance is an intrinsic property of
the system rather than due to imperfections in the sample's
border.\cite{comment}

\subsection{Phase diagram}
The transition line $H^*(T)$ from  avalanche to smooth flux
penetration regimes in Pb films with antidots was previously
determined from dc-magnetization and ac-susceptibility
measurements.\cite{hebert,silhanek} In the former case, the vortex
avalanche regime manifests itself as a jumpy response of the
magnetization, whereas in ac-susceptibility measurements the
signature of the transition between the different flux penetration
regimes is a local paramagnetic reentrance in the
ac-screening.\cite{silhanek}

In Fig.\,\ref{hstar} we plot the $H^*(T)$ lines previously
reported using ac-susceptibility together with the those
determined by ZFC MO measurements. Fig.\,\ref{hstar}(a) shows the
phase boundary obtained for samples with the same antidot array
and a slightly ($15\%$) different thickness. The remarkable
agreement between these two type of experiments reinforces the
interpretation of the reentrance in the ac-screening as the onset
of dendritic vortex avalanches. For comparison, the boundary lines
corresponding to samples with and without antidots are shown in
Fig.\,\ref{hstar}(b). In this case both samples were deposited
simultaneously and have the same thickness. The $H^*(T)$ line for
the non-patterned sample was determined by MO imaging whereas the
boundary for the antidot sample was obtained by ac-susceptibility
and is the same as already shown in Ref.[\onlinecite{silhanek}].
In Fig.\,\ref{hstar}(b) we can clearly see that the flux-jump
regime
 covers a larger portion of the phase diagram for the patterned sample than for the plain film, in
 agreement with previous reports.\cite{hebert,silhanek}

H\'{e}bert {\it et al.}\cite{hebert} proposed that the larger
extension of the avalanche regime in presence of antidots can be
related to the formation of a multi-terrace critical
state\cite{cooley,vvmprb} in this kind of samples. Within this
model, the main precursor of avalanches is the abrupt local change
$\delta B(x)$ in between terraces of constant $B$. However, the
direct observation of vortex dendrites indicates that this
scenario is not  appropriate to describe the observed extension of
the flux-jump regime.

\subsection{Dendrite field evolution}

In order to gain more insight into the dynamics of the avalanches
we studied the magnetic induction profile inside the dendrites in
ZFC experiments where the external field is increased in discrete
steps. In Fig.\,\ref{profiles}(a) a 3D image of the field
distribution  near one edge of the sample with antidots at $T = 4$
K and $\mu_0H= 1.8$ mT is given. Fig.\,\ref{profiles}(b) shows the
magnetic field profile inside the dendrite along the line
indicated by A-A' in Fig.\,\ref{profiles}(a) for different applied
fields. During the experiment the external field was increased in
steps of $\delta H=0.2$ mT but in Fig.\,\ref{profiles}(b) for
clarity we show only curves at $\delta H=0.4$ mT. In both figures
a maximum of $B$ at the edge of the sample is clearly seen as
expected for a thin film in a transverse magnetic field due to
demagnetization effects.

In general, we observe that once a dendrite develops, its {\it
shape} remains practically unchanged as the field is further
increased. Additionally, as  can be seen in
Fig.\,\ref{profiles}(b), the internal field $B$ along a dendrite
(A-A' line) increases as one moves from the edge towards the
center of the sample. Moreover, the magnitude of the magnetic
induction inside the dendrite can be even higher than the field at
the edge of the sample. After a dendrite has formed, the initial
deficiency of vortices near the edge of the sample is
progressively filled by new avalanches  as the external field is
ramped up (see for example the field profiles for $H \geq 2.4$ mT
in Fig.\,\ref{profiles}(b)). As already pointed out by Barkov {\it
et al.},\cite{shantsev} the initial inhomogeneous distribution of
vortices along the dendrites can be attributed to the field
induced by the screening currents that flow around the dendrite.
The  field lines associated with these currents are more dense
near  the front of the dendrite giving rise to higher local field
at that point.

Before the avalanche event occurs, the field penetrates following
a Bean-like profile ($H<1.2$ mT in Fig.\,\ref{profiles}(b)).
Interestingly, right after the avalanche develops, this slope
relaxes, as expected for a field-cooling process, and for higher
applied fields it recovers again. A similar effect was previously
observed in Nb films.\cite{welling}

The evolution of the  shape of the dendrites in their  transverse
direction is shown in Fig.\,\ref{profiles}(c). These profiles are
calculated along the line defined by B-B' in
Fig.\,\ref{profiles}(a) where no side branches of the dendrites
are crossed. For the sake of clarity the curves have been
displaced vertically. Naturally, the appearance of new peaks as
the field is increased corresponds to the formation of new
dendrites. The average width of the core of the dendrites is $w
\sim 45 \mu$m $\sim 30 d$, thus involving many unit cells of the
periodic pinning array. From this sequence of profiles it can be
seen that {\it the width of the peaks does not change with field}.
Also, no clear temperature dependence of the dendrite width has
been observed.

\subsection{Avalanche size distribution}

Since MO imaging maps the spatial distribution of $B$ inside the
sample we can calculate not only the total magnetic flux involved
in all avalanches but also the number of vortices involved in a
single avalanche.  In order to do that, we subtract two
consecutive MO images such that only relative changes are
recorded. Then we identify all avalanches that took place at a
given change in magnetic field and calculate the area, $A_i$, and
the magnetic flux, $\Phi_i$ involved in  each single avalanche
event. Also  we sum all the events for a given field step. The
resulting values $\Phi_T=\sum \Phi_i$ and $A_T=\sum A_i$ for all
fields and at three different temperatures are shown in
Fig.\,\ref{aval}(a) and (b).

Since avalanches stop at a temperature dependent field $H^*(T)$
the analysis  is significant up to a certain field that is smaller
for higher temperatures as can be seen from Fig.\,\ref{hstar}.  It
is interesting to note that the avalanches start at a field
$\mu_0H \sim 0.7$ mT (see   vertical dotted line in
Fig.\,\ref{aval}). This minimum magnetic field is independent of
the used magnetic field step and temperature, thus indicating that
it is a characteristic field of these type of instabilities. The
existence of a minimum field for the development of the first
avalanche or the first flux-jump, was predicted theoretically and
observed in many experiments. \cite{alsthuler} This feature was
also found in recent numerical and analytical studies of
instabilities of field penetration in thin films with random
disorder.\cite{aranson}

In Fig.\,\ref{aval}(a) we  observe a noisy behavior of $\Phi_T$ as
$H$ increases, in agreement with the observed jumpy
magnetization.\cite{hebert} Fig.\,\ref{aval}(b) shows the area of
the sample invaded by vortex avalanches for each step of $\mu_0H$.
Within the inherent noise due to avalanche behavior, the data in
both figures collapse onto a single curve for all temperatures.
From the data shown in these figures, we can roughly estimate the
internal field increment $\delta B$ at each external $\delta
H=0.4$ mT step. Indeed, from Fig.\,\ref{aval}(a) we have $\Phi_T
\sim 8 . 10^{-10}$ Tm$^2$ and from Fig.\,\ref{aval}(b), $A_T \sim
0.3$ mm$^2$, thus we obtain $\delta B = \Phi_T \times A_T \sim
2.6$ mT within the avalanches which is approximately 6 times
larger than the change in the applied field. This difference, of
course, is a consequence of the inhomogeneous distribution of
vortices in the avalanche regime leading to a strongly focussed
flux penetration. Averaging over the whole sample gives $\delta B
\sim \delta H$. The observation of a $\delta B$ independent of $H$
and $T$ is consistent with the previously reported temperature
independent flux-jumps in similar samples.\cite{hebert,silhanek}

We have pointed out in  Section III.A that dendrites at $T=5.5$ K
exhibit more branching than at lower temperatures (see
Fig.\,\ref{images}). This effect becomes evident in
Fig.\,\ref{aval}(c) where the average avalanche size, $<\Phi>$, is
plotted as a function of $\mu_0H$ for the same three temperatures
as in (a) and (b). The average is calculated for each applied
field as $<\Phi>=\sum \Phi_i / N$ where $N$ is the  number of
avalanches that takes place at that field. From the figure it is
evident that at $T=$3.5 K and 4.5 K the average avalanche sizes
are similar whereas at $T=$5.5 K there is a substantial increase
in $<\Phi>$. This result is consistent with a scenario where
finger-like dendrites of a well-defined size dominate at low $T$
whereas large, highly branched tree-like dendrites, with no
characteristic size, dominate at high $T$. The present analysis of
avalanche sizes shows that even though the average size of the
dendrites depends on temperature, the total flux involved in all
avalanches remains approximately constant.

In addition to the analysis of magnetic flux and area of
avalanches presented above, similar to a bulk magnetization
measurement, the identification of each avalanche event allows us
to analyze the {\it distribution} of individual avalanche sizes in
the whole dendritic penetration regime. Presently, there is
controversy on whether the critical state in superconductors is a
self organized critical (SOC) system or not. \cite{radovan}  In
the first case the size distribution will be described by a power
law since avalanches of all sizes are expected.  In our samples we
find that the avalanche size distribution (we define the size by
the total amount of moved flux) is consistent with power law
behavior for $T < 5.5$ K (see Fig.\,\ref{counts}). However, at
large avalanche sizes the data departs from a linear behavior in
the log-log scale. This is due to a finite size effect since the
length of the dendrites is limited by the size of the
sample\cite{authors2}. Clearly we observe power-law behavior over
one and a half decade consistently with SOC behavior. At $T=5.5$ K
a reliable fitting is not possible since the small amount of
avalanches results in a very poor statistics (in this case there
are of the order of 80 avalanches while at 4.5 K the number is
200). The power-law exponent extracted from the fitting of the
data at low temperatures is $\tau \sim 0.9(1)$. A similar value
($\tau =1.09$) has been obtained recently by Radovan and
Zieve\cite{radovan} in Pb plain films by analyzing the size of the
magnetization jumps using local Hall probe measurements. For
YBa$_2$Cu$_3$O$_7$ Aegerter {\it et al.} \cite{wellingepl} found a
slightly larger value $\tau =1.29(2)$.

\section{Conclusions}

We have studied magnetic flux penetration in Pb thin films with
antidots by means of MO imaging. At low temperatures and fields
the penetration is dominated by vortex avalanches while at higher
$T$ and $H$ a rather smooth  and flat flux front is observed. We
have found that the avalanches develop in the form of dendrites
similarly to previous observations in Nb films with a periodic
antidot array. The morphology of the dendrites  changes with
temperature, from finger-like at low $T$ to tree-like at high $T$.
For all $H-T$ we observe that the vortex motion is guided by the
pinning potential generated by the antidots. In general, new
dendrites are formed far from old dendrites, in regions where
previously no invasion of vortices has taken place, indicating
that they interact repulsively. This occurs until there is no room
for a new dendrite. As a consequence, the emergence of new
dendrites leads to a more uniform magnetic field distribution.

The boundary between dendritic and smooth penetration as
determined by MO imaging  is in a very good agreement with the
results obtained from ac-susceptibility measurements in similar
samples, see Fig.\,\ref{hstar}. We have also corroborated that in
the film with antidots the vortex avalanche regime is extended to
higher temperatures and fields as compared to the case of
unpatterned films. The detection of dendritic penetration in the
flux-jump regime shows that the proposed model
\cite{hebert,silhanek} of multi-terrace formation for  flux
penetration is not applicable at low temperatures for Pb films
with antidots. Magnetic field profiles inside the dendrites
indicate that the field at the tip of the dendrite is of the order
or even higher than the field at the edge of the sample. This is
due to the high field induced by the screening currents, which
make a hairpin bend at the end of the dendrite. A relaxation of
the magnetic field slope at the edge of the sample due to the
avalanche is observed.

Avalanche size distribution analysis shows that the sum of the
flux over all avalanches remains constant with temperature. This
accounts for the observed temperature independent magnetization at
low $H$ and $T$. However, we find that the average size of the
dendrites depends on temperature. Thus, a detailed knowledge of
the morphology of avalanches is necessary for a complete
description of flux penetration in these superconducting thin
films. Besides this global analysis of vortex avalanches we study
the size distribution of individual avalanches taking profit of
the local character of our technique. We find that the size
distribution of individual avalanches  is consistent with a power
law behavior over more than a decade of avalanche sizes at low
temperatures. However, the absence of finite size scaling analysis
\cite{authors2} does not allow to make a definite conclusion on
whether the system is SOC or not.

 Recently, it was demonstrated that the coupling
between nonlocal flux diffusion with local thermal diffusion can
account for dendritic penetration in plain films. We believe that
a similar analysis for the case of samples with periodic pinning
will be very helpful for a complete understanding of magnetic flux
instabilities in superconducting samples.

 \acknowledgments
We would like to thank R. Jonckheere for fabrication of the resist
patterns. This work was supported by the Belgian Interuniversity
Attraction Poles (IUAP), Research Fund K.U.Leuven GOA/2004/02, the
Fund for Scientific Research Flanders (FWO) and ESF ``VORTEX''
program and by FOM (Stichting voor Fundamenteel Onderzoek der
Materie) which is financially supported by NWO (Nederlandse
Organisatie voor  Wetenschappelijk Onderzoek).

\bibliographystyle{prsty}

\begin{thebibliography}{10}
\bibitem{comment} Small imperfections as indentations in
the sample's border lead to a deformation of the screening
currents that can trigger or promote dendrite nucleation. In our
particular case one of the substrate edges was deliberately cut to
create a rough sample edge, however no clear evidence of a higher
dendrite density was detected.
\bibitem{authors2} We did not perform a finite size scale
analysis because the anisotropic shape of avalanches  (elongated
in one direction) introduces artifacts in the size distribution
when selecting a system length, L, smaller than the size of the
sample.
\end{thebibliography}

\newpage

\section{Figure Captions}

\begin{figure}[hhh]
\caption{MO images of the Pb sample AD75 with antidots showing
different types of flux penetration: (a) finger-like dendritic
penetration at  $\mu_0H = 1.2$ mT and $T =4.5$ K  and (b) $\mu_0H
= 1.2$ mT and $T =5.5$ K , (c) tree-like dendritic outburst
coexisting with smooth flux penetration at  $\mu_0H = 1.5$ mT and
$T =6$ K, and  (d) smooth profile at $\mu_0H = 1.5$ mT and $T
=6.5$ K. A saw-tooth-like magnetic wall domain artifact from the
magneto-optical garnet is observed in the bottom part of the
images. (e) and (f) MO images of a Pb plain film at $T = 2.5$ K
for $\mu_0H = 0.4$ mT and $\mu_0H =1.2$ mT  respectively. The
scale bar in each figure corresponds to 0.5 mm.}
  \label{images}
\end{figure}

\begin{figure}[hhh]
\centering
\caption{ Phase boundary lines, $H^*(T)$, separating dendritic
from smooth penetration for different samples. Open symbols
correspond to ac-susceptibility measurements \cite{silhanek} while
filled symbols are values obtained by MO imaging. (a) Results for
samples with antidots AD65 and AD75. (b) Comparison between a
plain film (PF15) and a sample with antidots (AD15) (see Table
I).} \label{hstar}
\end{figure}

\begin{figure}[hhh]
\centering
\caption{(a) Three-dimensional plot of magnetic field  near the
edge of the sample (AD75) with antidots  at $T= 4$ K and
$\mu_0H=1.8$ mT. (b) Profiles of magnetic field  along a dendrite
indicated by A-A' in (a) for different applied fields.  (c)
Sequence of magnetic field profiles along the line B-B' in (a)
(transverse to the dendrites). For clarity the curves are
displaced by 1.5 mT along the vertical axis. The difference in
external field between each consecutive profile is 0.2 mT. }
\label{profiles}
\end{figure}

\begin{figure}[hhh]
  \centering
\caption{Avalanches as a function of applied field for sample
AD75. (a) Total magnetic flux, $\Phi_T$, (b) area of the sample
covered by avalanches, $A_T$ and (c) average size of avalanches,
$<\Phi>$, at each field step as a function of $H$ for different
temperatures. }
  \label{aval}
\end{figure}

\begin{figure}[hhh]
  \centering
\caption{Log-log plot of the distribution of avalanche sizes for
sample AD75 at different temperatures. The size of an avalanche is
given by the moved flux $\Phi$. The straight solid lines are
linear fits, yielding a power-law exponent $\tau = 0.9(1)$.}
  \label{counts}
\end{figure}

\end{document}